\documentclass[iop]{emulateapj}

\usepackage{rotating}

\shorttitle{A Pulsar Wind Nebula in the Cygnus Loop} 
\shortauthors{Katsuda et al.}

\begin{document}

\title{Discovery of a Pulsar Wind Nebula Candidate in the Cygnus Loop}

\author{Satoru Katsuda\altaffilmark{1}, Hiroshi Tsunemi\altaffilmark{2},
Koji Mori\altaffilmark{3}, Hiroyuki Uchida\altaffilmark{4},
Robert Petre\altaffilmark{5}, Shin'ya Yamada\altaffilmark{1}, and 
Toru Tamagawa\altaffilmark{1}
}

\altaffiltext{1}{RIKEN (The Institute of Physical and Chemical
  Research), 2-1 Hirosawa, Wako, Saitama 351-0198}

\altaffiltext{2}{Department of Earth and Space Science, Graduate School
of Science, Osaka University, 1-1 Machikaneyama, Toyonaka, Osaka,
60-0043, Japan}

\altaffiltext{3}{Department of Applied Physics, Faculty of Engineering,
University of Miyazaki, 1-1 Gakuen Kibana-dai Nishi, Miyazaki, 889-2192,
Japan}

\altaffiltext{4}{Department of Physics, Kyoto University, 
Kitashirakawa-oiwake-cho, Sakyo, Kyoto 606-8502, Japan}

\altaffiltext{5}{NASA Goddard Space Flight Center, Code 662, Greenbelt
MD 20771}




\begin{abstract}
We report on a discovery of a diffuse nebula containing a pointlike 
source in the southern blowout region of the Cygnus Loop supernova 
remnant, based on {\it Suzaku} and {\it XMM-Newton} observations.  
The X-ray spectra from the nebula and the pointlike source are well 
represented by an absorbed power-law model with photon indices 
of 2.2$\pm$0.1 and 1.6$\pm$0.2, respectively.  
The photon indices as well as the flux ratio of
$F_\mathrm{nebula}$/$F_\mathrm{pointlike}\sim4$ lead us to propose that 
the system is a pulsar wind nebula, although pulsations have not yet 
been detected.  If we attribute its origin to the Cygnus Loop supernova, then
the 0.5--8\,keV luminosity of the nebula is computed to be 
2.1$\times10^{31}$\,($d$/540\,pc)$^{2}$\,ergs\,s$^{-1}$, where $d$ is 
the distance to the Loop.  This implies a spin-down loss-energy 
$\dot{E}\sim2.6\times10^{35}$\,($d$/540\,pc)$^{2}$\,ergs\,s$^{-1}$.  
The location of the neutron star candidate, $\sim$2$^{\circ}$ away 
from the geometric center of the Loop, implies a high transverse velocity 
of $\sim$1850\,($\theta$/2$^{\circ}$)\,($d$/540\,pc)\,($t$/10\,kyr)$^{-1}$\,km\,s$^{-1}$, assuming the currently accepted age of the Cygnus Loop.
\end{abstract}

\section{Introduction}

The Cygnus Loop supernova remnant (SNR), the X-ray image of which is 
shown in Fig.\ref{fig:RASS_image}, is one of the brightest SNRs in the 
X-ray sky.  Because of its proximity, $540 (+100, -80) $\,pc 
\citep{Blair2009}, it is an extremely important 
object that is often called a prototypical middle-aged SNR 
\citep[10\,kyr:][]{Levenson1998}.  The X-ray morphology is an almost 
perfect circular shell except for a southern blowout region.  The origin 
of the blowout has been a matter of debate; it may be caused by either low ambient density \citep{Aschenbach1999,Uchida2008} or a second SNR 
\citep{Uyaniker2002,Sun2006}.

Another mystery for the Cygnus Loop is the absence of a central compact 
remnant.  It is
believed that the Cygnus Loop is the result of a core-collapse SN,
because the blast wave is now hitting the walls of the cavity that was most
likely created by a strong stellar wind from the progenitor
\citep[e.g.,][]{Charles1985,Hester1994,Levenson1997}.
The comparatively small size of the cavity ($R\sim13$\,pc at a distance 
of 540\,pc) led \citet{Levenson1998} to suggest that the progenitor 
star was of spectral type later than B0, $\sim$15M$_\odot$.  This view is 
further supported by recent X-ray abundance measurements which have led to 
progenitor mass estimates of 12--15M$_\odot$ 
\citep[e.g.,][]{Tsunemi2007,Kimura2009,Uchida2011}.  
Such a progenitor star should have formed a neutron star during the SN 
explosion.  Although considerable effort has been devoted to searching 
for a neutron star in the Cygnus Loop over nearly three decades, none has 
been found yet \citep[e.g.,][]{Miyata1998,Miyata2001}.

Here, we report on the discovery of a possible pulsar wind nebula 
(PWN) in the southern blowout region of the Cygnus Loop, based on 
X-ray observations with {\it Suzaku} and {\it XMM-Newton}.  This 
Letter focuses on the discussion about the PWN candidate, while other results 
based on these data have been published elsewhere 
\citep[e.g.,][]{Tsunemi2007,Uchida2011,Katsuda2011}.

\section{Observations}

\begin{figure}[htbt]
\begin{center}
\includegraphics[angle=0,scale=1.72]{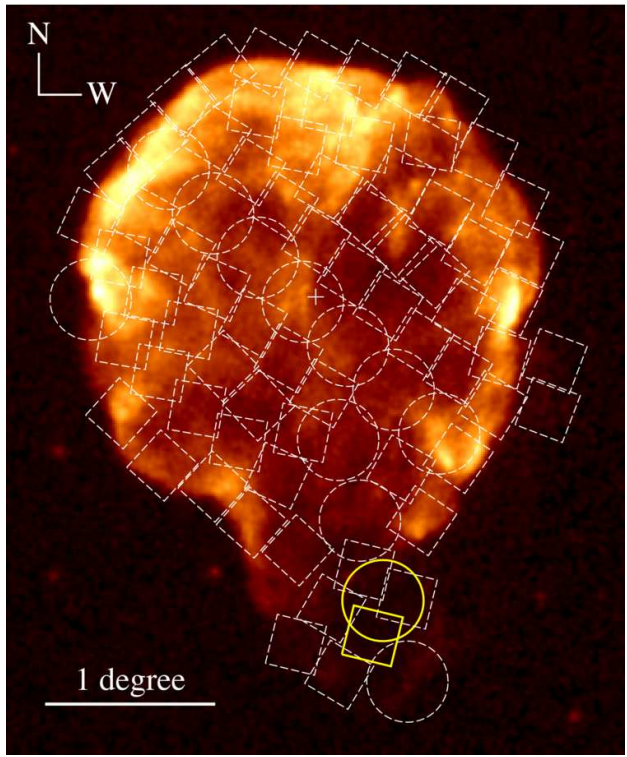}\hspace{1cm}
\caption{{\it ROSAT} all-sky survey image of the Cygnus Loop, scaled as the 
square root of the surface brightness.  The energy band used is 0.2--2\,keV.  
All of the {\it Suzaku} XIS and the {\it XMM-Newton} 
EPIC fields observed to date are overlaid as boxes and circles, respectively.
The XIS and EPIC images of the yellow FOV is shown in 
Fig.~\ref{fig:XIS_XMM_image}.  The geometric center of the Cygnus Loop 
is indicated by a cross.
} 
\label{fig:RASS_image}
\end{center}
\end{figure}

We have conducted over 80 pointing observations of the Cygnus Loop 
using the {\it Suzaku} X-ray Imaging Spectrometer \citep[XIS:][]{Koyama2007} 
and the {\it XMM-Newton} European Photon Imaging Camera 
\citep[EPIC:][]{Turner2001,Struder2001}, covering nearly the entirety of 
this large SNR.  The fields of view (FOV) of these observations are overlaid on the {\it ROSAT} 
all-sky survey image in Fig.~\ref{fig:RASS_image}.  We focus here on one
XIS/EPIC observation located in the southern blowout region.
The XIS and EPIC observations were performed on 2011-05-07 (Obs.ID: 
506013010) and 2006-05-15 (Obs.ID: 0405490301), respectively.

For the {\it Suzaku} XIS data, we use cleaned event data prepared by the 
{\it Suzaku} operations team.  The net exposure time is 60.3\,ks for each 
XIS.  The {\it XMM-Newton} data suffer severely from high background (BG)
flares due to soft protons throughout the observation.  Nonetheless, the
EPIC image may provide us with useful information on spatial structures,
thanks to much better angular resolution than the XIS.  We thus use 
relatively clean time regions, where the count rates in the 5--12\,keV 
are less than 5\,cts\,s$^{-1}$ for MOS1/2 or 50\,cts\,s$^{-1}$ 
for pn.  The effective exposure times obtained are 7.8\,ks, 8.0\,ks, and 
6.0\,ks for MOS1, MOS2, and pn, respectively.  All the raw data are 
processed using version 11.0.0 of the XMM Science Analysis Software.  
Further analyses are done with {\tt heasoft} tools of version 6.11.1 and 
the latest CALDB files updated on 2011-11-09.

\section{Analysis and Results}

\begin{figure*}
\begin{center}
\includegraphics[angle=0,scale=2.2]{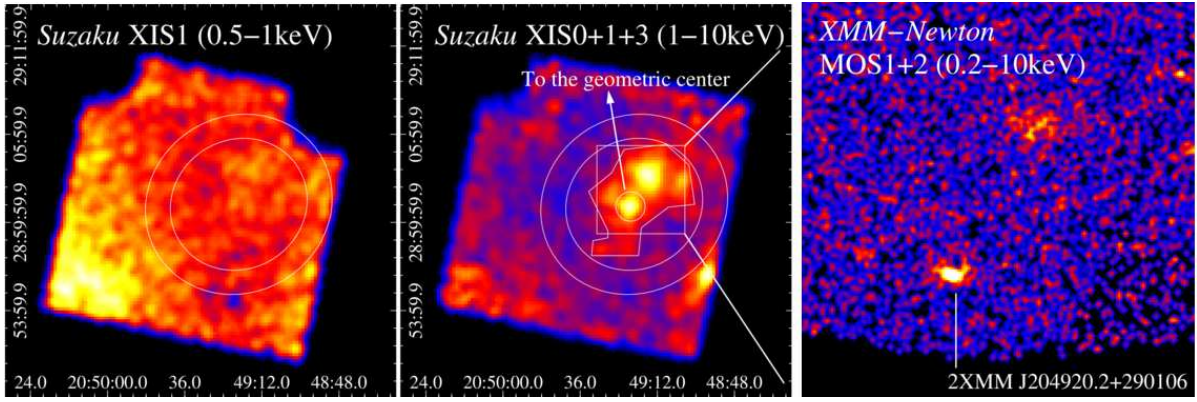}\hspace{1cm}
\caption{Left: Soft-band (0.5--1\,keV) XIS1 image of the yellow box in 
Fig.~\ref{fig:RASS_image}.  The image is scaled as a square root of the 
surface brightness and corrected for vignetting effects after subtraction 
of non X-ray BG.  
A white elliptical annulus shows where we extract a local BG.
Center: Same as left but in the hard band.  The data taken by XIS0, XIS1, 
and XIS3 are summed to improve photon statistics.  The spectral 
extraction regions are shown as a white circle (for the pointlike source) 
and a polygon (for the nebula).  
Right: Closeup image of the white box (6$^{\prime}$ square) in 
Fig.~\ref{fig:XIS_XMM_image} center taken by the {\it XMM-Newton} MOS1+2.  
BG is not subtracted and vignetting effects are not corrected.  
} 
\label{fig:XIS_XMM_image}
\end{center}
\end{figure*}

Through our comprehensive analyses of the X-ray data, we have discovered a 
hard X-ray--emitting nebula in the southern blowout region of the Cygnus 
Loop, i.e., the XIS/EPIC FOV illustrated in yellow in 
Fig.~\ref{fig:RASS_image}.  Figure~\ref{fig:XIS_XMM_image} left and center 
are vignetting-corrected XIS images in 0.5--1\,keV and 1--10\,keV, 
respectively, for which non X-ray BG is subtracted by using the 
{\tt xisnxbgen} software \citep{Tawa2008}.  While the left panel is 
dominated by thermal emission from the Cygnus Loop, the central panel 
represents mixture of cosmic X-ray BG and hard X-rays from astrophysical 
sources.  In the hard-band image, we see a bimodal diffuse feature---our 
target of interest in this Letter.  An {\it XMM-Newton} image in 
Fig.~\ref{fig:XIS_XMM_image} right successfully resolves it into a 
northern diffuse nebula and a southern pointlike source.  It should be 
noted that the short exposure time and the presence of soft proton 
events in the {\it XMM-Newton} image make it difficult to detect faint 
diffuse emission between the pointlike source and the northern nebula.

The position of the pointlike source is determined to be [RA, Dec] = 
[20:49:20.309, +29:01:05.57 (J2000)], by using the {\tt emldetect} 
software.  The statistical uncertainty is negligible compared with the
astrometric uncertainty of 2$^{\prime\prime}$ \citep[based on the
{\it XMM-Newton} Calibration Technical Note -- ][]{Guainazzi2011}.  
The source has been identified as 2XMM~J204920.2+290106 in the 
{\it XMM-Newton} Serendipitous Source Catalog \citep{Watson2009}.  
No obvious optical and infrared counterpart is found in the Palomar 
Observatory Sky Survey and 2MASS All Sky Survey, respectively, although 
there is an apparent compact galaxy whose peak is $\sim$3$^{\prime\prime}$ 
away from the X-ray pointlike source and its optical emission extends to 
the source position.  The R-band upper limit is estimated to be 
R$\gtrsim$19 mag.  No radio counterpart is detected in an 
archival VLA 1.4\,GHz image.  The radial profile of the pointlike 
source is found to be consistent with the EPIC point-spread function at 
the same off-axis angle.  The northern nebula is positionally consistent 
with an infrared object 2MASX J20491447+2903237 which is identified as 
an extended extra galactic source \citep{Strutskie2006}, while its extent 
appears to be quite small and thus there is no evidence for the physical 
association to the X-ray nebula.


We examine the XIS spectra of the pointlike source and the surrounding 
nebula.  The spectral extraction regions, a 1$^{\prime}$-radius circle 
around the pointlike source and a polygon tracing the edge of the diffuse 
nebula, are shown in Fig.~\ref{fig:XIS_XMM_image} center.  BG is taken 
from an elliptical annulus shown in Fig.~\ref{fig:XIS_XMM_image} 
left and center. The BG produced in this way includes not only usual X-ray BG 
but also thermal emission from the Cygnus Loop itself 
\citep[typically, $kT_\mathrm{e}\sim$0.3\,keV and 
$n_\mathrm{e}t\sim$10$^{11}$\,cm$^{-3}$\,s:][]{Uchida2008}.

\begin{figure*}
\begin{center}
\includegraphics[angle=0,scale=0.6]{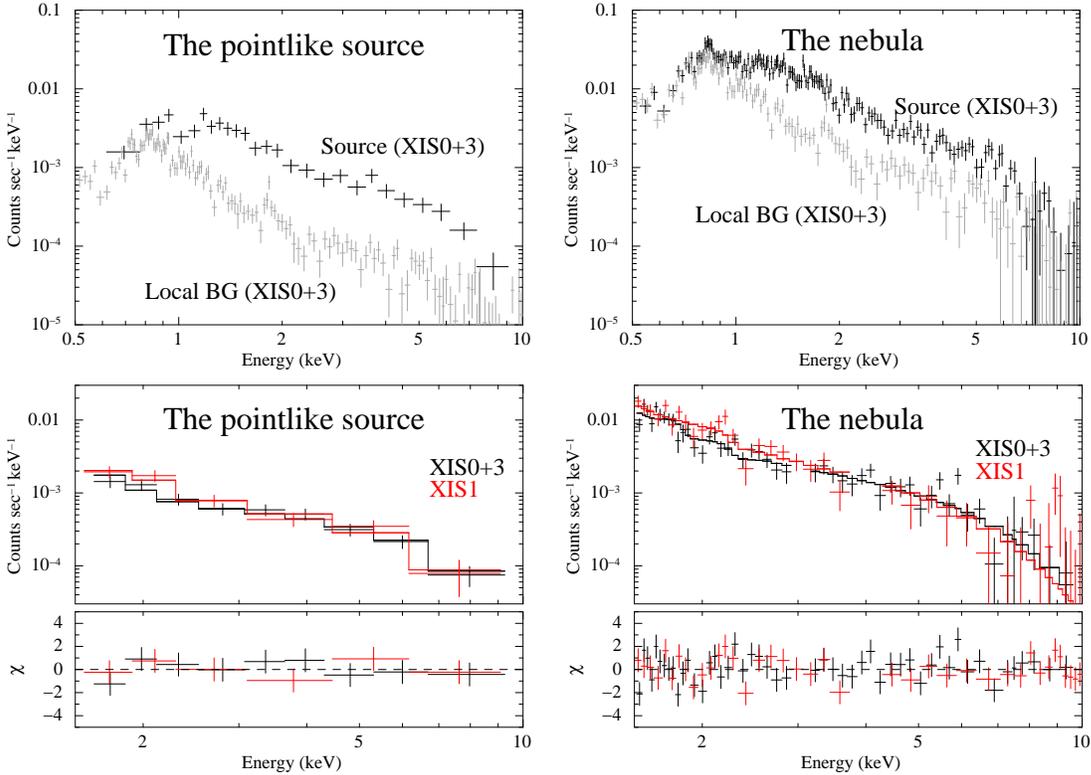}\hspace{1cm}
\caption{Left upper: {\it Suzaku} XIS (0+3) spectrum for the pointlike 
source (black) and for the area-normalized BG spectrum (gray).
Right upper: Same as left but for the diffuse nebula.
Left lower: Local-BG subtracted XIS spectra for the pointlike source 
along with the best-fit power-law model.  Black and red correspond to 
the FI (XIS0+3) and the BI (XIS1), respectively.  Lower panel shows 
residuals.
Right lower: Same as left but for the diffuse nebula.
} 
\label{fig:spec}
\end{center}
\end{figure*}

The XIS0+3 spectra together with the area-normalized local-BG spectra 
are shown in the upper panels of Fig.~\ref{fig:spec}, and the local-BG 
subtracted spectra are plotted in the lower panels.
The photon numbers after BG subtraction are 753 and 4534 for the 
pointlike source and the nebula, respectively.  Since emission below 
1.5\,keV is dominated by Cygnus Loop's thermal emission, it is 
in principle difficult to subtract the local BG properly in the 
soft X-ray band.  We thus use the 1.5--10\,keV band for  
spectral analysis.  In this case, the intervening column density cannot 
be constrained, so we tentatively fix the value of $N_\mathrm{H}$ 
to 4$\times$10$^{20}$\,cm$^{-2}$ which is typical for the Cygnus Loop 
\citep[e.g.,][]{Inoue1980}.  
Before fitting, each spectrum is grouped into bins with at least 
$\sim$50 counts in prior to background subtraction, which allows us 
to perform a $\chi^2$ test. In the calculation of $\chi^2$ values, we 
use the standard weighting method, i.e., the square root of detected 
counts.

We find that either a power-law model or an optically thin thermal 
emission model \citep[{\tt apec}:][]{Smith2001} can adequately describe 
the data.  The thermal emission model, however, requires a high electron 
temperature of 2--4\,keV, and very low abundances of $\lesssim$0.8 times 
the solar value.  Such a high temperature has never been reported from 
the Cygnus Loop \citep[e.g., ][]{Tsunemi2007}, and the low abundances are 
unusual in astrophysical sources.  We thus reject the thermal model 
from an astrophysical point of view.
Also, a blackbody model is safely rejected from a statistical point of 
view.  Adding a blackbody or thermal emission component to the power-law 
component does not improve the fits, either.  In this way, we conclude that 
the power-law is the most reasonable model to describe the data.  
The best-fit parameters 
from the power-law model are summarized in Table~\ref{tab:param}.  
We note that consistent results are obtained if we let the soft X-ray band 
(i.e., 0.5--1.5\,keV) remain and fit the spectra by a power-law plus local-BG 
model whose normalization is allowed to vary freely.  Also, no significant 
spatial variation of the photon index in the synchrotron nebula has 
been detected owing to insufficient photon statistics.

\begin{deluxetable}{lccc}
\tabletypesize{\tiny}
\tablecaption{Power-law spectral-fit parameters}
\tablewidth{0pt}
\tablehead{
\colhead{Region}&\colhead{$\Gamma$}&\colhead{Flux$^{a}$}&\colhead{$\chi^{2}$/d.o.f.}}
\startdata
The pointlike source & 1.6$\pm$0.2  & 2.0$\pm$0.5 & 6.6/12 \\
The nebula & 2.2$\pm$0.1 & 5.5$^{+0.7}_{-0.6}$ & 96.1/86
\enddata
\tablecomments{Errors quoted are at 90\% confidence level.\\
$^a$Unabsorbed 0.5--8\,keV flux in units of 
10$^{-13}$\,ergs\,s$^{-1}$\,cm$^{-2}$.}
\label{tab:param}
\end{deluxetable}

To search for long-term spectral variability of the pointlike source, we 
also fit the EPIC spectra.  While the data are affected by soft protons, 
flux estimates would be relatively robust if we subtract a local BG.  
We find that the fluxes (and photon indices) inferred from the MOS and pn 
spectra are marginally consistent with those inferred using the XIS.  
In addition, a Fourier search does not give significant pulsations.
This might be due to  poor photon statistics as well as insufficient time 
resolution (0.073\,s, 2.6\,s, and 8\,s for pn, MOS, and XIS, respectively).


\section{Discussion}

We have presented the discovery of a diffuse nebula in the Cygnus Loop 
containing a pointlike source.  The spectra of both are well 
represented by a power-law model, reminding us of a PWN.  A PWN origin is 
further supported by the following observational features.  
\citet{Kargaltsev2008} reported a strong correlation between PWN 
luminosities and the nonthermal luminosities of pulsars, namely 
$L_\mathrm{PWN\,(0.5-8\,keV)}$/$L_\mathrm{PSR\,(0.5-8\,keV)}$ is nearly 
constant at 4.  Our measured 
$L_\mathrm{nebula\,(0.5-8\,keV)}$/$L_\mathrm{pointlike\,(0.5-8\,keV)}$ 
of $\sim$3.8 is in good agreement with this correlation.  Above, we 
assume that the nebula 
extends over the pulsar candidate with its mean surface brightness, 
resulting in 
$F_\mathrm{nebula\,(0.5-8\,keV)}\sim5.7\times$10$^{-13}$\,ergs\,cm$^{-2}$\,s$^{-1}$ and 
$F_\mathrm{pointlike\,(0.5-8\,keV)}\sim1.5\times$10$^{-13}$\,ergs\,cm$^{-2}$\,s$^{-1}$.  
Furthermore, the measured photon indices are consistent with those of 
PWNe, and the nebula's photon index is too hard to be 
interpreted as shell emission from relativistic particles accelerated at 
an SNR shock (generally, $\Gamma\sim2.5-3$).  It should be also noted that
the spectrum of the nebula is softer than the pointlike source, typical 
of PWNe.  From a morphological point of view, the brightest part of the 
nebula is offset from the pointlike source.  
This is unusual for PWNe, but complicated structures are often 
found in PWNe \citep[e.g.,][]{Kargaltsev2008}; hence the morphology cannot 
be a strong reason to eliminate the PWN hypothesis.
Also, the pointlike source is not likely to be a background active 
galactic nucleus or a foreground normal star, since it has no obvious 
radio or optical counterpart and its X-ray spectrum is too hard for 
a normal star.  While low mass X-ray binaries (LMXBs) 
have high X-ray to optical luminosity ratios, this possibility is ruled 
out from the low X-ray flux of the pointlike source,
$F_\mathrm{2-10\,keV}\sim$1.3$\times$10$^{-13}$\,ergs\,cm$^{-2}$\,s$^{-1}$ 
based on the XIS; a normal LMXB luminosity of 
$\sim$10$^{36}$\,ergs\,s$^{-1}$ requires a distance of 250\,kpc which is 
far outside the Galaxy.  On the other hand, these properties are consistent 
with the expectations for an isolated neutron star, even though the 
conclusive evidence of a pulse period is lacking.  Nevertheless, we 
propose that this system is a PWN

We note, however, another possibility: that the system consists of a galaxy 
cluster and an unknown pointlike source which might not be related to the 
cluster.  A brief justification can be made by comparing the nebula's 
properties with those of galaxy clusters.  A $L_X$-temperature correlation 
of galaxy clusters \citep[e.g.,][]{Fukazawa2004} would result in nebula's 
luminosity, $L_\mathrm{2-10\,keV} = 
2.0\times10^{44}$\,($kT$/3\,keV)$^{2.79}$\,ergs\,s$^{-1}$.
Combining the luminosity with nebula's flux, $F_\mathrm{2-10\,keV} = 
2.7\times10^{-13}$\,ergs\,s$^{-1}$, we obtain a distance of $\sim$2.5\,Gpc.  
The apparent radius of 3$^{\prime}$ can be then translated to a real 
radius of $\sim$2\,($d$/2.5\,Gpc)\,Mpc, which is typical for galaxy clusters 
with $kT$ of 3\,keV \citep{Fukazawa2004}.  Therefore, the X-ray 
luminosity, temperature, and size are all in agreement with those of galaxy 
clusters.  In addition, based on the {\it ROSAT} PSPC survery of galaxy 
clusters \citep{Burenin2007}, a surface density of clusters is $\sim$0.3 
deg$^{-2}$ at $\sim$2.8$\times10^{-13}$ ergs\,cm$^{-2}$\,s$^{-1}$ which 
is the nebula's brightness in 0.5-2 keV.  Given the Cygnus Loop's large 
dimension of $\sim$10 degree$^{2}$, we expect $\sim$3 clusters inside the 
Loop.  Thus, a chance probability to detect a galaxy cluster in the 
Loop is fairly large.

While other possibilities have not been excluded, this system is most
likely a PWN based on its observed properties.  
We thus further discuss the PWN scenario.  Whether it is physically 
associated with the Cygnus Loop is an important point.  Evidence favoring 
a common origin is the fact that, even though the Cygnus Loop 
should have originated from a core-collapse SN, our survey of the 
{\it Suzaku} and {\it XMM-Newton} data has not detected any PWN or 
neutron star except for the PWN candidate disclosed here.  If the objects
are associated, then the distance to the PWN should be the same as the 
Cygnus Loop, i.e., $d\sim$540\,pc \citep{Blair2005}.  If this is the case, 
the 0.5--8\,keV luminosities of the pulsar candidate and the nebula are 
computed to be 5$\times10^{30}$ ($d$/540\,pc)$^{2}$\,ergs\,s$^{-1}$ and 
2.1$\times10^{31}$ ($d$/540\,pc)$^{2}$\,ergs\,s$^{-1}$, respectively.  
This would be one of the lowest luminosities measured from the 
$\sim$50 known X-ray PWNe \citep{Kargaltsev2008}.  The physical size of 
the nebula would be 
$\sim$0.5\,($\theta_\mathrm{r}$/3$^{\prime}$)($d$/540\,pc)\,pc, 
where $\theta_\mathrm{r}$ is the angular radius of the nebula.  This size is an 
order of magnitude smaller than expected based on the PWN size-age 
relation reported by \citet{Bamba2010}.  This discrepancy is mitigated
if the PWN is farther away than the Cygnus Loop.  Therefore, the 
PWN candidate is either a distant, ordinary PWN unrelated to the Cygnus 
Loop or a nearby, peculiar PWN.  This point should be revisited by 
future detailed X-ray spectroscopy that will allow for an $N_\mathrm{H}$ 
measurement.

We next estimate some important parameters, as is usually performed 
for putative PWNe \citep[e.g.,][]{Hughes2001,Olbert2003,Gaensler2003}.  
A well-known relation between PWN luminosity and spin-down 
loss power, $\eta \equiv L_X/\dot{E}\sim8\times10^{-5}$ 
\citep[most recently,][]{Vink2011}, indicates that 
$\dot{E}\sim2.6\times10^{35}$\,($d$/540\,pc)$^{2}$\,ergs\,s$^{-1}$.  
The estimated $\dot{E}$ allows us to deduce a pulsar period of 
$P \sim 0.48\,[2/(n-1)]^{0.5}\,(\dot{E}/2.6\times10^{35}\,\mathrm{ergs\,s^{-1}})^{-0.5}\,(t/10\,\mathrm{kyr})^{-0.5}$\,s, assuming a standard moment of 
inertia ($I = 10^{45}$\,g\,cm$^{2}$, i.e., a standard neutron star radius 
of 10\,km and a mass of 1.4\,M$_\odot$), a braking index of 3, 
and a negligible initial period.  Also, the period derivative can be 
estimated to be 
$\dot{P} = P/[(n-1)t] \sim 7.7\times10^{-13}\,(\dot{E}/2.6\times10^{35}\,\mathrm{ergs\,s^{-1}})^{-0.5}\,(t/10\,\mathrm{kyr})^{-1.5}$\,s\,s$^{-1}$, 
leading to a surface magnetic field estimate of $B \sim 1.9\times10^{13} 
(\dot{E}/3.4\times10^{35}\,\mathrm{ergs\,s^{-1}})^{-0.5} 
(t/10\,\mathrm{kyr})^{-1}$\,G.  
Comparing these parameters with those of the other X-ray--emitting PWNe 
\citep{Kargaltsev2008}, we find that $\dot{E}$ is somewhat 
smaller than others derived at the age of 10\,kyr and that the $P$ and $B$ 
values are among the largest of all the listed PWNe.  Alternatively,
noting that there is a correlation between $\dot{E}$ and $\tau$, 
$\dot{E} = 41.1 - 1.08$\,log\,$\tau$ for $\tau < 10^{4}$\,kyr, we can infer 
$\tau$ for the estimated $\dot{E}$ of 
$2.6\times10^{35}$\,($d$/540\,pc)$^{2}$\,ergs\,s$^{-1}$ to be $\sim$180\,kyr.  
This would lead to $P\sim112$\,ms and $B\sim10^{12}$\,G, which are common 
for X-ray PWNe \citep{Kargaltsev2008}.  We also point out that the estimated 
$\dot{E}$ implies a gamma-ray luminosity for a possible pulsar of 
1.6$\times$10$^{34}$\,($\dot{E}/2.6\times10^{35}\,\mathrm{ergs\,s^{-1}})^{-0.5}$\,ergs\,s$^{-1}$ in 0.1--100\,GeV \citep{Abdo2010}.  This should be easily 
detected with the {\it Fermi} LAT, which is not the case 
\citep{Katagiri2011}.

It is interesting to note that the neutron star candidate is located far 
from Cygnus Loop's geometric center defined by {\it Einstein} 
\citep{Ku1984}, as can be seen in Fig.~\ref{fig:RASS_image}.  If we 
assume that the pointlike source started moving from the geometric 
center 10\,kyr ago, its transverse velocity is 
$\sim$1850\,($\theta$/2$^{\circ}$)($d$/540\,pc)($t$/10\,kyr)$^{-1}$\,km\,s$^{-1}$,
and its proper motion
$\sim$0$^{\prime\prime}$.72\,($\theta$/2$^\circ$)($t$/10\,kyr)$^{-1}$\,yr$^{-1}$,
where $\theta$ is the angular distance between the neutron star candidate and 
the geometric center and $t$ is the best-estimated age of the Cygnus Loop.
Such a high velocity is one of the fastest of known neutron stars 
\citep[e.g.,][]{Kaspi2006}.  However, we should keep in mind that the 
velocity estimate is quite uncertain.  
The explosion location is quite uncertain (and thus 
$\theta$), because the SN went off in a cavity.  In fact,
it has been proposed that the SN occurred in the southern blowout 
region \citep{Tenorio-Tagle1985}.  The age is also a matter of debate; 
it might not be accurate to 50\%.

We have checked that the existing data including {\it Einstein}, {\it 
ROSAT}, and {\it ASCA} do not allow us to measure the proper motion of the 
source; the source is not detected, possibly due to insufficient exposure 
times as well as low detection efficiency in the hard X-ray band.  
It is also difficult to measure proper motions using the XIS data, given 
the large position uncertainty of $\sim$1$^{\prime}$ due to thermal 
fluctuation \citep{Uchiyama2008}.
Thus, high-resolution images with {\it Chandra} or {\it XMM-Newton} will be 
required.  Confirmation of a large proper motion towards the south would be 
direct evidence that the pointlike source is indeed the neutron star remnant of 
the Cygnus Loop.  In this case, we expect a cometary bowshock structure 
around the neutron star \citep[cf.][]{Gaensler2004} due to the high 
Mach number of $\sim$6, given a reasonable sound speed of 
$c_\mathrm{s} = \sqrt{(\gamma kT)/(\mu m_\mathrm{p})}\sim300$\,km\,s$^{-1}$, 
where $\gamma$ is a specific heat ratio (5/3), $kT$ is the plasma 
temperature (0.3\,keV), and $\mu$ is the mean molecular mass ($\sim$0.6 
for solar abundance plasmas), and $m_\mathrm{p}$ is the proton mass.  
Note that such a bow shock would be directed away from the geometric 
center of the Loop and thus not necessarily related to the northern 
nebula.  A much higher Mach number would be expected if the system is 
already outside the hot plasma and is now proceeding into cold interstellar 
medium.  On the other hand, if the proper motion turns out to be small, the 
location of the PWN candidate near the center of the blowout appears to 
support the suggestion that the southern blowout is a separate SNR 
\citep{Uyaniker2002}.


\section{Conclusion}

Using {\it Suzaku} and {\it XMM-Newton}, we discovered a hard 
X-ray--emitting diffuse nebula containing a pointlike source in the 
southern blowout region of the Cygnus Loop.  Their properties suggest 
that it is most likely a PWN.  The physical relation
to the Cygnus Loop is still uncertain at this point.  Future 
observations are essential in order to confirm that the object is a PWN as well 
as to clarify whether it is associated with the Cygnus Loop.

\acknowledgments

We would like to thank Drs. Teruaki Enoto and Takao Kitaguchi for
fruitful discussions.
S.K.\ and S.Y.\ are supported by the Special Postdoctoral Researchers 
Program in RIKEN.  This work is partly supported by a Grant-in-Aid for 
Scientific Research by the Ministry of Education, Culture, Sports, 
Science and Technology (23000004).


\end{document}